\documentclass{article}

\usepackage{amsmath,amsfonts,amssymb,times,graphicx,natbib,algorithm,algorithmic,hyperref}

\usepackage[accepted]{data4good2016}

\icmltitlerunning{Predicting Zoonotic Tick Vectors}

\begin{document}

\twocolumn[
\icmltitle{Predicting Novel Tick Vectors of Zoonotic Disease}

\icmlauthor{Barbara A. Han}{hanb@caryinstitute.org}
\icmladdress{Cary Institute of Ecosystem Studies,
            Box AB, Millbrook, NY 12545 USA}
\icmlauthor{Laura Yang}{ylaura9797@gmail.com}
\icmladdress{Spackenkill High School,
            112 Spackenkill Rd, Poughkeepsie, NY 12603 USA}

\vskip 0.3in
]

\begin{abstract}
With the resurgence of tick-borne diseases such as Lyme disease and the emergence of new pathogens such as Powassan virus, understanding what distinguishes vector from non-vector species, and predicting undiscovered tick vectors is an important step towards mitigating human disease risk. We apply generalized boosted regression to interrogate over 90 features for over 240 species of Ixodes ticks. Our model predicted vector status with ~97\% accuracy and implicated 14 tick species whose intrinsic trait profiles confer high probabilities (~80\%) that they are capable of transmitting infections from animal hosts to humans. Distinguishing characteristics of zoonotic tick vectors include several anatomical structures that facilitate efficient host seeking and blood-feeding from a wide variety of host species. Boosted regression analysis produced both actionable predictions to guide ongoing surveillance as well as testable hypotheses about the biological underpinnings of vectorial capacity across tick species.
\end{abstract}

\section{Introduction}
\label{sec:intro}

Ticks transmit a greater diversity of pathogenic agents than any other arthropod \citep{morand_taxonomy_2006} and are responsible for vectoring at least 30 zoonotic infectious diseases worldwide \citep{gideon_global_1994}. With global warming, tick-borne diseases are projected to increase even more drastically \citep{levi_accelerated_2015}. Unsurprisingly, a large volume of research is dedicated to understanding tick biology, with substantial bias towards a fraction of tick species known to vector pathogens from animals to humans (zoonotic vectors). Among the hard-bodied ticks (Family Ixodidae), the most species-rich genus, Ixodes, contains 244 species, of which 34 are known zoonotic vectors. What enables these few species to acquire and transmit zoonotic disease? Identifying which features distinguish effective zoonotic vectors from non-vector species is essential for understanding what drives vectorial capacity, and for developing preemptive approaches to reducing tick-borne diseases in humans. 
We applied a machine learning method called generalized boosted regression \citep{elith_working_2008,ridgeway_gbm:_2013} to identify which features best predict tick species capable of transmitting zoonotic diseases. This machine learning algorithm determines which features are most important for correctly predicting a response variable (here, a binary variable designating whether the tick species is a zoonotic vector) by building thousands of linked trees that successively improve upon the predictions of the previous tree. This model-free approach does not rely on distributional assumptions and is ideal for high-dimensional ecological data containing hidden, nonlinear interactions, and non-random patterns of missing data \citep{han_rodent_2015,di_marco_human_2015}. Our goal was to determine which tick species might harbor undiscovered zoonotic pathogens, and to identify traits of Ixodes ticks that best predict their status as vectors of human zoonotic disease.

\section{Methods}
\subsection{Data Collection}

Based on nomenclature from a standard reference \citep{guglielmone_hard_2014}, we searched published literature for species binomials of 244 ticks of genus Ixodes. The response variable for our analysis was a binary score assigned to each species (0 or 1) based on their zoonotic vector status as established by the GIDEON database \citep{berger_gideon:_2005}, which we used to provide the public health consensus on the status of each tick species as vectors for at least one zoonotic disease. From peer-reviewed primary literature, we collated a total of 104 traits across three life stages (larvae, nymph, and adult) per tick species. These traits can be partitioned into four categories: anatomy, biology, geography, and pathology. All anatomical features were standardized to millimeters.

\subsection{Analysis}

Using similar approaches applied successfully in previous studies \citep{han_rodent_2015,han_undiscovered_????}, we applied generalized boosted regression via the gbm package \citep{ridgeway_gbm:_2013} in R \citep{r_core_team_r:_2014}. We tuned the model to build an ensemble of 30,000 trees using 10-fold cross-validation with a shrinkage rate of 0.00025 and an interaction depth of 3. Boosted regression models accommodate missing data by treating “missingness” as a value by using surrogate splits \citep{ridgeway_gbm:_2013}, which draws from the correlation structure among trait variables. However, we also set a minimum threshold of 1\% data coverage across tick species as criteria for inclusion in the model in order to remove those variables with extremely low coverage. The difference between using all traits and removing those with less 1\% coverage were negligible to model performance. Data were randomly partitioned into training (70\%) and test data (30\%). Prediction accuracy was measured by AUC.

\subsection{Study Bias}

While many epidemiological metrics (e.g., prevalence of tick-borne disease in humans) are biased by study effort (i.e., the amount of healthcare or research spending per country), traits describing intrinsic vector biology (e.g., body size, clutch size) are less subject to the same biases. 

However, if variation in study effort across tick species leads to significant differences in data coverage (i.e., biological features are only known for vectors), this type of bias can affect model results. To diagnose this possible issue, we produce a plot showing the probability of a tick species being a novel vector as a function of citation count (Figure 1).

\begin{figure}[t]
\begin{center}
\centerline{\includegraphics[width=\columnwidth]{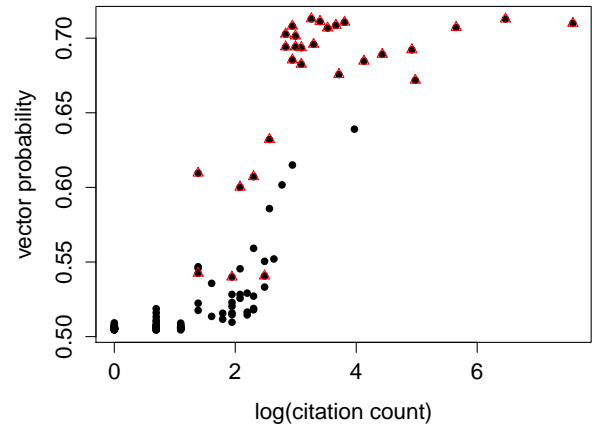}}
\caption{The study effort recorded as the number of studies in Web of Science reporting each tick species’ Latin binomial (log(citation count)).}
\label{fig:graph}
\end{center}
\vskip -0.2in
\end{figure} 

\section{Results}
\label{sec:results}

Our model was able to distinguish vector from non-vector species with 97.8\% accuracy. The best predictors of zoonotic vector status included host breadth (number of orders and families that tick feeds on); tarsus I length of larvae; capitulum lengths of larvae, nymph, and female adult stages; female adult scutum length; clutch size; and female body length (Figure 1). 

Compared to non-vectors, tick species that vector zoonotic disease tend to have several distinctive characteristics.  These zoonotic vectors have wider host breadth, feeding on host species from five or more families and four or more orders. In addition, the larvae possess shorter tarsus I lengths (the length of the first segment of the first pair of legs; \textless0.18mm), whereas those of non-vectors are generally longer. Larvae and adult female ticks have shorter capitulum lengths than non-vectors, but nymphs exhibit the opposite pattern, with nymphs of known zoonotic vectors having longer capitula than non-vectors. Species that are known zoonotic vectors have capitulum lengths of 1.00mm, 0.125mm, and 0.400-0.800mm for adult females, larvae, and nymphs, respectively. Lastly, the adult females have larger clutches (\textgreater1000 eggs), greater scutum length (\textgreater1.0mm), and longer bodies, both while engorged (\textgreater6.0mm) and unengorged (\textgreater2.5mm) compared to adult females of non-vector species. 

Of 244 Ixodes species, 34 species are currently recognized as vectors of human diseases (Berger, 2005). Our model identifies 14 additional potential zoonotic vectors, which share similar trait profiles with those 34 species. Among these 14 predicted vectors, 10 species are recognized by primary literature as possibly parasitizing humans (Table 1). The remaining 4 species, \emph{I. canisuga}, \emph{I. trichosuri}, \emph{I. eldaricus}, and \emph{I. aragaoi}, are novel vectors that have not yet been identified as human parasites but reflect >80\% probability of vectoring a zoonotic pathogen.

We found intrinsic traits poorly predicted which tick species that were well-studied. Although tick species that are known vectors for at least one zoonosis have higher citation counts, there are also several tick species (vectors and non-vectors) that are reasonably well studied with both low and high probabilities of being zoonotic vectors. Taken together, these results confirm that the tick trait profile reported here reflects that of a zoonotic vector rather than that of well studied tick.


\begin{table}[ht]
\caption{Model predictions of novel zoonotic tick vectors, ranked in order of descending probability, and journal articles postulating human infestation for each species}
\label{table:numbers}
\vskip 0.15in
\begin{center}
\begin{small}
\begin{sc}
\setlength{\tabcolsep}{2pt}
\begin{tabular}{|c|c|c|}
\hline
Rank         & Species       & Human Infestation \\\hline
1            & $I. rubicundus$         & \citep{horak_ixodid_2002}         \\\hline
2            & $I. canisuga$        & $None$        \\\hline
3            & $I. acuminatus$         & \citep{hillyard_ticks_1996}         \\\hline
4            & $I. vespertilionis$        & \citep{piksa_first_2013}        \\\hline
5            & $I. sculptus$        & \citep{salkeld_host_2006}        \\\hline
6            & $I. apronophorus$         & \citep{fedorov_ixodoidea_1967}         \\\hline
7            & $I. woodi$         & \citep{merten_state-by-state_2000}         \\\hline
8            & $I. kingi$         & \citep{salkeld_host_2006}         \\\hline
9            & $I. kazakstani$          & \citep{filippova_ixodid_1977}         \\\hline
10           & $I. redikorzevi$         & \citep{emchuk_certain_1968}         \\\hline
11           & $I. trichosuri$         & $None$         \\\hline
12           & $I. eldaricus$         & $None$         \\\hline
13           & $I. laguri$         & \citep{bursali_review_2012}         \\\hline
14           & $I. aragaoi$         & $None$         \\\hline
\end{tabular}
\end{sc}
\end{small}
\end{center}
\vskip -0.1in
\end{table}

\section{Discussion}
\label{sec:discussion}

Predicting tick vectors for future zoonotic diseases is a critical step toward disease prevention and will rely on understanding what features enable ticks to be better human disease vectors. Here, we report a profile of tick traits that distinguish zoonotic vectors from non-vectors with more than 90\% accuracy, and we identify several tick species with high probabilities of vectoring one or more zoonotic diseases as potential targets for increased investigation and surveillance. 

We found that the most important predictor of zoonotic vector status in Ixodes ticks was the diversity of vertebrate species that the tick parasitizes. This finding is consistent with the general principle that the probability of vectoring a zoonotic disease correlates directly with host breadth \citep{davies_phylogeny_2008,woolhouse_host_2005}. This pattern has been widely postulated for vector groups such as mosquitoes, and we find evidence that this is also true for Ixodes ticks. 

Several anatomical features were highly predictive of vector status. Larvae of vector species tend to have shorter tarsus I lengths (length of the first segment of the first pair of legs) compared to non-vectors. The larval stage is important because acquisition of zoonotic pathogens (e.g., Lyme spirochetes) often occurs during the blood meal at this stage \citep{matuschka_stage-associated_1992}. Moreover, if infected at this stage, larvae have two potentially infectious bites through which to transmit the pathogen compared to one bite if it is infected as a nymph. Tarsus I contains many important sensory organs that promote important behaviors across all 3 life stages, including seeking ideal habitat, host-seeking, and mate-seeking behaviors. In addition, tarsus I contains Haller’s organ, a vital organ for determining host location, host odors, detecting pheromones, and serving other environmental sensory functions \citep{sonenshine_biology_2013}. We are unaware of studies examining allometric scaling patterns of tarsus I and/or Haller’s organ, but from these results we spectulate that if the length of tarsus I is correlated to host-seeking at the larval stage, shorter tarsus I may cause larvae to be less selective about hosts or environments, thereby causing vectors at this life to be less selective about where and from whom the first blood meal is acquired. Similarly, if Haller’s organ correlates directly with the length of tarsus I, larvae with shorter tarsus I lengths would be less selective about vertebrate host species. Reduced host selectivity could lead to more generalized feeding preferences across a wide diversity of host types, increasing the possibility of contact with hosts infected with zoonotic pathogens.

\begin{figure}[t]
\begin{center}
\centerline{\includegraphics[width=\columnwidth]{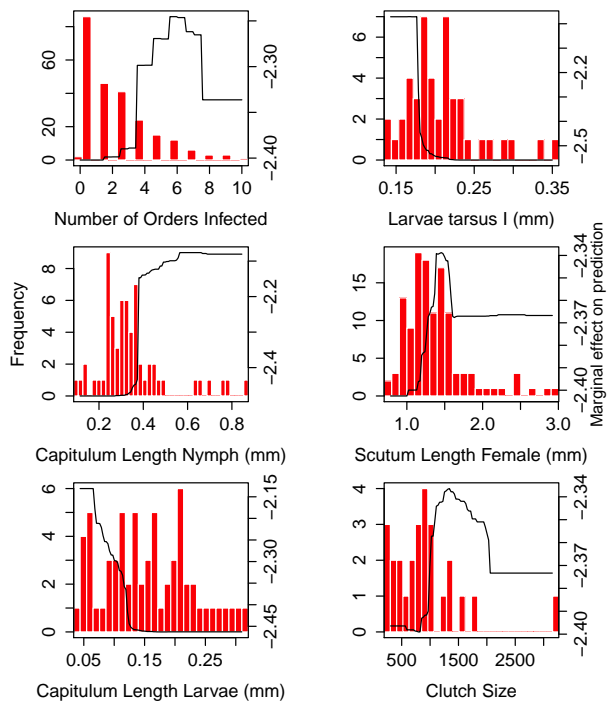}}
\caption{Partial dependence plots representing the most important predictor variables from a generalized boosted regression model predicting tick zoonotic vector status. These partial dependence plots illustrate how ticks that transmit zoonotic diseases have distinctive characteristics from non-vectors. In each plot, the frequency histogram represents the trait values for all tick species (vectors and non-vectors) (left y axis), and the black line shows the trait values for tick vectors (right y-axis). For example, while the majority of Ixodes ticks parasitize hosts from few taxonomic orders (histogram bars, panel 1), tick vectors of zoonotic diseases tend to feed upon hosts from 4 or more orders (line).}
\label{fig:graph}
\end{center}
\vskip -0.2in
\end{figure} 

Another important trait distinguishing vectors from non-vectors was the capitulum length in larvae, nymphs, and adult females. Capitulum length is determined by hypostome length and salivarium size in ticks. The hypostome is a the ratchet-like anchor within the capitulum that is inserted into the host body \citep{richter_how_2013}, and the salivarium is a repository that collects and delivers tick saliva.  Tick saliva contains bioactive molecules responsible for facilitating blood meals as well as zoonotic pathogens such as \emph{Borrelia burgdorferi} (causative agent of Lyme disease), \emph{Francisella tularensis} (causative agent of tularemia), and others \citep{reuben_kaufman_ticks:_2010}. We found that capitulum lengths were shorter in adult female and larval vectors than those of non-vectors, and that capitulum length in nymphal vectors was longer in zoonotic vectors. This pattern is consistent with widely documented patterns of vector competence of Ixodes species that transmit pathogens that cause anaplasmosis, babesiosis, and Lyme disease: of the three developmental stages, the nymphal stage is disproportionately responsible for human transmission \citep{matuschka_stage-associated_1992}. With softer substrates like those encountered in human and other mammal hosts, ticks benefit from a more secure anchor conferred by deeper penetration of mouthparts that comprise the capitulum \citep{richter_how_2013}. Secure attachments lead to increased feeding times, which increase the probability of successful transmission for tick-borne diseases such as Lyme disease \citep{kazimirova_tick_2013}. Thus, capitulum length at the nymphal stage may be a reliable indicator of the vectorial capacity of Ixodes tick species for zoonotic pathogens. 

Our analysis also revealed that tick vectors have a fecundity advantage over non-vector ticks \citep{shine_evolution_1988}. This profile supports the fecundity advantage model which asserts that larger females produce larger clutches. Specifically, body size, scutum length, and clutch size of adult females are all larger for zoonotic vectors compared to non-vectors. Larger body sizes enable the ingestion of larger blood meals from hosts, leading to greater available resources for egg production \citep{ford_relationships_1989}. Combined, the trait profile produced by our analysis shows that zoonotic tick vectors are most likely to be species where adult females produce a larger number of eggs which develop into larvae that may be feeding on a greater diversity of species. These larvae develop into nymphs whose capitula allow for more efficient and longer feeding times on soft-bodied hosts compared to non-vector species, which produces larger adult females with greater fecundity. 

In addition to identifying a profile distinguishing tick species which are zoonotic vectors from non-vectors, our model identified 14 Ixodes tick species with ~80\% probabilities of being undiscovered vectors of zoonotic disease (Table 1). The majority of these species reside in Nearctic or Palearctic biomes, and all of them are habituated to forest or grassland habitats \citep{guglielmone_hard_2014}. Some of the ticks we identify are suspected in the primary literature as being likely disease vectors, but these species are not currently recognized by the public health community as zoonotic vectors \emph{per se}. For example, one species, \emph{Ixodes acuminatus}, is capable of transmitting \emph{Borrelia burgdoferi} sensu lato, though it is not considered an important vector for human disease in nature, perhaps due to low contact with human populations \citep{morse_factors_1995}. The saliva of another species, \emph{Ixodes rubicundus}, can cause paralysis in sheep \citep{fourie_seasonal_1989}, but there is no record of this species transmitting infectious disease to humans. Given that many of the 14 predicted vector species are understudied and some of them are already suspected to have potential health consequences for humans (Table 1), our study offers new utility for identifying tick species whose intrinsic traits suggest they should be surveillance targets. In addition to informing the biological basis by which some ticks vector zoonotic pathogens, our study underscores the crucial importance of basic research on ticks and other arthropod vectors, since understanding the biological basis of transmission ecology will rely fundamentally on understanding the innate characteristics distinguishing vector from non-vector species.

\bibliography{Yang_and_Han_Tick_Vectors_ms}
\bibliographystyle{icml2016}

\end{document}